\documentclass[aps,prb,twocolumn,groupedaddress,superscriptaddress,nobibnotes]{revtex4-1}
\usepackage{amsmath}
\usepackage{amssymb}
\usepackage{graphicx}
\usepackage{dcolumn}
\usepackage{bm}
\usepackage{color}
\usepackage{subfig}
\usepackage[normalem]{ulem}
\usepackage[none]{hyphenat}
\usepackage{siunitx}
\usepackage{float}

\newcommand{\MST}{MnSb$_2$Te$_4$}
\newcommand{\MBT}{MnBi$_2$Te$_4$}
\newcommand{\MBTBT}{MnBi$_4$Te$_7$}
\newcommand{\MBTBTBT}{MnBi$_6$Te$_{10}$}
\newcommand{\MSTST}{MnSb$_4$Te$_7$}
\newcommand{\MSTSTST}{MnSb$_6$Te$_{10}$}

\newcommand{\ST}{Sb$_2$Te$_3$}
\newcommand{\BT}{Bi$_2$Te$_3$}

\begin{document}

\title{Topological magnetic materials of the (MnSb$_2$Te$_4$)$\cdot$(Sb$_2$Te$_3$)$_n$ van der Waals compounds family}

\author{S.\,V. Eremeev}
\email{eremeev@ispms.tsc.ru}
\affiliation{Institute of Strength Physics and Materials Science, Russian Academy of Sciences, 634021 Tomsk, Russia}
\affiliation{Tomsk State University, 634050 Tomsk, Russia}

\author{I.\,P. Rusinov}
\affiliation{Tomsk State University, 634050 Tomsk, Russia}

\author{Yu.\,M.~Koroteev}
\affiliation{Institute of Strength Physics and Materials Science, Russian Academy of Sciences, 634021 Tomsk, Russia}
\affiliation{Tomsk State University, 634050 Tomsk, Russia}

\author{A.\,Yu. Vyazovskaya}
\affiliation{Tomsk State University, 634050 Tomsk, Russia}
\affiliation{Saint Petersburg State University, 198504 Saint Petersburg, Russia}

\author{M.\, Hoffmann}
\affiliation{Institut f\"ur Theoretische Physik, Johannes Kepler Universit\"at, A 4040 Linz, Austria}

\author{P.\,M. Echenique}
\affiliation{Departamento de F\'{\i}sica de Materiales UPV/EHU, 20080 Donostia-San Sebasti\'{a}n, Basque Country, Spain}
\affiliation{Donostia International Physics Center (DIPC), 20018 Donostia-San Sebasti\'{a}n, Basque Country, Spain}
\affiliation{Centro de F\'{i}sica de Materiales (CFM-MPC), Centro Mixto CSIC-UPV/EHU,  20018 Donostia-San Sebasti\'{a}n, Basque Country, Spain}

\author{A.~Ernst}
\affiliation{Institut f\"ur Theoretische Physik, Johannes Kepler Universit\"at, A 4040 Linz, Austria}
\affiliation{Max-Planck-Institut f\"ur Mikrostrukturphysik, Weinberg 2, D-06120 Halle, Germany}

\author{M.\,M. Otrokov}
\email{mikhail.otrokov@gmail.com}
\affiliation{Centro de F\'{i}sica de Materiales (CFM-MPC), Centro Mixto CSIC-UPV/EHU,  20018 Donostia-San Sebasti\'{a}n, Basque Country, Spain}
\affiliation{IKERBASQUE, Basque Foundation for Science, 48011, Bilbao, Spain}

\author{E.\,V. Chulkov}
\email{evguenivladimirovich.tchoulkov@ehu.eus}
\affiliation{Departamento de F\'{\i}sica de Materiales UPV/EHU, 20080 Donostia-San Sebasti\'{a}n, Basque Country, Spain}
\affiliation{Donostia International Physics Center (DIPC), 20018 Donostia-San Sebasti\'{a}n, Basque Country, Spain}
\affiliation{Centro de F\'{i}sica de Materiales (CFM-MPC), Centro Mixto CSIC-UPV/EHU,  20018 Donostia-San Sebasti\'{a}n, Basque Country, Spain}
\affiliation{Saint Petersburg State University, 198504 Saint Petersburg, Russia}

\date{\today}

\begin{abstract}
Combining robust magnetism, strong spin-orbit coupling and  unique thickness-dependent properties of van der Waals crystals could enable new spintronics applications. Here, using density functional theory, we propose the (MnSb$_2$Te$_4$)$\cdot$(Sb$_2$Te$_3$)$_n$ family of stoichiometric van der Waals compounds that harbour multiple topologically-nontrivial magnetic phases. In the groundstate, the first three members of the family, i.e. \MST\, ($n=0$), \MSTST\, ($n=1$), and \MSTSTST\, ($n=2$), are 3D antiferromagnetic topological insulators (AFMTIs),
while for $n \geq 3$ a special phase is formed, in which a nontrivial topological order coexists with a partial magnetic disorder in the system of the decoupled 2D ferromagnets, whose magnetizations point randomly along the third direction. Furthermore, due to a weak interlayer exchange coupling, these materials can be field-driven into the FM Weyl semimetal ($n=0$) or FM axion insulator states ($n \geq 1$). Finally, in two dimensions we reveal these systems to show intrinsic quantum anomalous Hall and AFM axion insulator states, as well as quantum Hall state, achieved under external magnetic field, but without Landau levels. Our results provide a solid computational proof that \MST\, is not topologically trivial as was previously believed that opens possibilities of realization of a wealth of topologically-nontrivial states in the (MnSb$_2$Te$_4$)$\cdot$(Sb$_2$Te$_3$)$_n$ family.
\end{abstract}

\maketitle

\section*{Introduction}
The recently discovered first intrinsic antiferromagnetic
topological insulator (AFMTI) \MBT\, has attracted a great deal of attention \cite{Otrokov.Nature,Otrokov.prl2019,Li.sciadv2019,Zhang.prl2019, Gong_CPL2019,Otrokov.2dmat2017,Otrokov.jetpl2017,Eremeev.jac2017,Eremeev.nl2018,Hirahara.nl2017, Hagmann.njp2017, Chowdhury.npjcmat2019, Lee.prr2019, Rienks.nat2019, Aliev.jac2019, Yan.prm2019, Zeugner.cm2019, Chen.ncomms2019,Vidal.prb2019, Deng.Sci2020, Yan.PRB2019,Liu.NatMat2020,Hu.NatComm2020,Wu.sciadv2019,Peng.prb2019, Lian.PRL2020,Hao.prx2019,Chen.PRX2019,Li.prx2019,Estyunin.aplmat2020,Swatek.PRB2020,Yan.PRL2020, Shikin.srep2020, Xu.arxiv2020,
Hirahara.ncomms2020}. The AFMTI state of matter is expected to give rise to a number of exotic phenomena such as quantized magnetoelectric effect \cite{Mong.prb2010}, axion electrodynamics \cite{Li.nphys2010}, and Majorana hinge modes \cite{Peng.prb2019}. Due to its layered van der Waals structure and interlayer AFM order, in the 2D limit \MBT\, has been predicted to show an unusual set of thickness-dependent magnetic and topological transitions, which drive it through FM and (un)compensated AFM phases, as well as quantum anomalous Hall (QAH) and zero plateau QAH states \cite{Otrokov.prl2019, Li.sciadv2019}. This makes \MBT\, the first stoichiometric material predicted to realize the zero plateau QAH state intrinsically, which has been theoretically shown to host the exotic axion insulator (AXI) phase \cite{Wang.prb2016}. Soon after the density functional theory (DFT) prediction  \cite{Otrokov.prl2019, Li.sciadv2019}, the evidence of both the AXI and QAHI states has indeed been reported in thin \MBT\, flakes in Refs. \cite{Liu.NatMat2020} and \cite{Deng.Sci2020}, respectively. Very recently, the QAH regime has also been reported in the \MBT/\BT\,
superlattices\cite{Deng.nphys2020}. Moreover, the quantized Hall effect under external magnetic field has been achieved in \MBT\, flakes, displaying the Chern numbers $C=1$ \cite{Deng.Sci2020,Liu.NatMat2020,Ge.NSR2020} and 2 \cite{Ge.NSR2020}. A recent study also predicts that the $C=3$ state is achievable in the twisted \MBT\, bilayer \cite{Lian.PRL2020}. Finally, \MBT-derived family of compounds \cite{Aliev.jac2019, Klimovskikh_npjQM2020} has been recently proposed as a magnetically tunable platform for realizing various symmetry-protected higher-order typologies \cite{Zhang.PRL2020, Roy.prb2020}, hinged quantum spin Hall effect \cite{Ding.prb2020}, helical Chern insulator \cite{Liu.arxiv2020}, as well as spin-polarized flat
bands \cite{Petrov.arxiv2020}.

The abovementioned results establish a new direction in the field of magnetic TIs that focuses on intrinsically magnetic stoichiometric compounds. For further successful development of this direction more layered magnetic materials need to be found. Currently, the quest for new \MBT-like systems follows the strategy of either codoping Bi sublattice with Sb \cite{Chen.ncomms2019, Yan.PRB2019,
Chen.PRM2020, Lee.arxiv2020} or growing the homologous series (\MBT)$\cdot$(\BT)$_n$ whose typical representatives are
\MBTBT\, ($n=1$) and \MBTBTBT\, ($n=2$) \cite{Aliev.jac2019, Hu.NatComm2020, Wu.sciadv2019, Jahangirli.jvst2019, Yan.prm2020, Klimovskikh_npjQM2020}, although MTIs with higher $n$ have also been reported very recently \cite{Jahangirli.jvst2019, Klimovskikh_npjQM2020, Hu.arxiv2019.2, Deng.nphys2020}.

In this paper, using the state-of-the-art DFT and tight-binding calculations we propose the (\MST)$\cdot$(\ST)$_n$ homologous series, which has not been considered previously. By means of the highly-accurate total-energy calculations we show that $n=0,1,$ and 2 systems, i.e. \MST, \MSTST, and \MSTSTST, are interlayer antiferromagnets in which FM Mn layers are coupled antiparallel to each other and the easy axis of staggered magnetization points perpendicular to the layers. Such magnetic ordering makes these compounds invariant with respect to the combination of the time-reversal ($\Theta$) and primitive-lattice translation ($T_{1/2}$) symmetries, $S=\Theta T_{1/2}$, which gives rise to the $\mathbb{Z}_2$ topological classification of AFM insulators \cite{Mong.prb2010,Fang.prb2013}, $\mathbb{Z}_2$ being equal to 1 for all these materials. Consistently, their $S$-breaking (0001) surfaces exhibit Dirac point gap,
which is a consequence of the out-of-plane direction of the easy axis of staggered magnetization. At the same time, the $S$-preserving surfaces are gapless as expected for AFMTIs. For the MnSb$_8$Te$_{13}$ compound ($n=3$), whose interlayer exchange coupling is so weak that it can hardly be reliably established within DFT, we predict the AFMTI state assuming the interlayer AFM phase. In the FM phases, that can be achieved by the application of the external magnetic field, \MST\, appears to be a Weyl semimetal, while \MSTST, \MSTSTST, MnSb$_8$Te$_{13}$, etc., are 3D FM AXIs according to the $\mathbb{Z}_4$ classification. Finally, we study the thin films of the (\MST)$\cdot$(\ST)$_n$ compounds and find AFM AXIs in the zero plateau QAH state, Chern insulators in the constrained FM state, as well as intrinsic QAHI in the uncompensated AFM state. Our study provides a compelling first-principles evidence that \MST, previously predicted to be trivial at normal conditions \cite{Zhang.prl2019, Chen.ncomms2019, Zhou.prb2020, Lei.arxiv2020, Liu.arxiv2020mst}, is, in fact, a topological insulator or Weyl semimetal in the AFM or FM states, respectively. These findings significantly extend the emergent family of the MTIs by predicting a number of
solid materials candidates.

\begin{figure*}[!bth]
    \begin{center}
    \includegraphics[width=\textwidth]{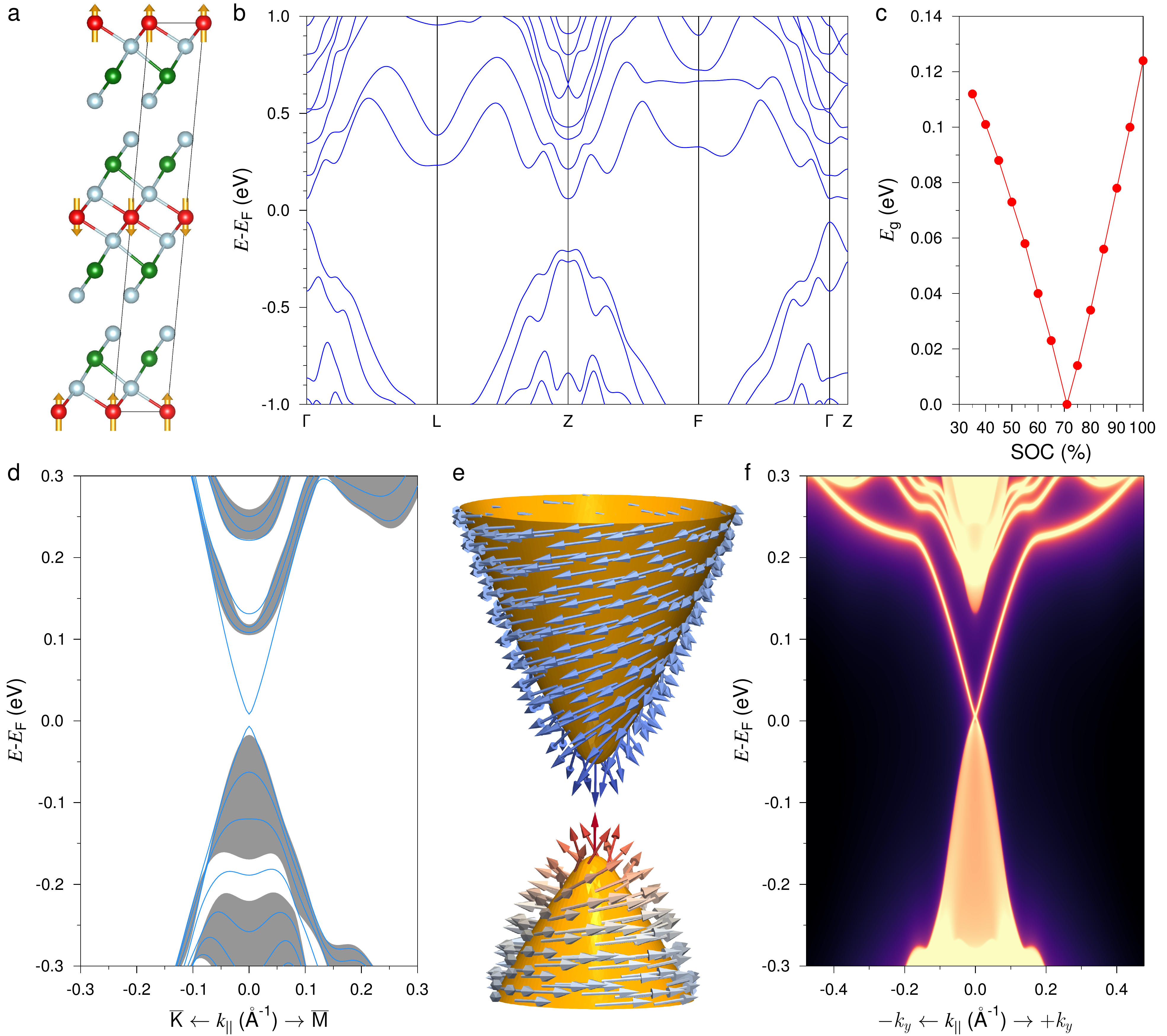}
    \end{center}
 \caption{{\bf Bulk and surface band structures of AFM \MST.} (a) Atomic structure of bulk \MST\ with red, gray and green balls showing Mn, Te and Sb atoms. (b) Bulk electronic band structure calculated for the interlayer AFM state modelled in the rhombohedral unit cell. (c) The evolution of the bulk band gap $E_g$ with the change of the SOC constant $\lambda$. (d) Electronic structure of the MnSb$_2$Te$_4$(0001) surface. Gray areas correspond to the bulk band structure projected onto the surface Brillouin zone (BZ). (e) Spin texture of the gapped Dirac state in the vicinity of the BZ center. (f) The tight-binding calculated electronic bandstructure of the $S$-preserving ($10 \bar 1 1$) surface. The regions with a continuous spectrum correspond to the 3D bulk states projected onto a 2D BZ.}
 \label{fig:124}
\end{figure*}

\section*{Results}

The growth of the trigonal $R\bar 3m$-group bulk MnSb$_2$Te$_4$ has been reported very recently \cite{Yan.PRB2019, Murakami.prb2019}. This structure, comprised of the septuple layer blocks (SLs; Fig.~\ref{fig:124}a) turns out to be  energetically more favorable than the monoclinic structure \cite{Eremeev.jac2017}, previously observed for some other chalcogenides with the same formula \cite{Ranmohotti.jacs2012}. Magnetic properties of \MST\, have been found to depend on the degree of Mn-Sb intermixing: the larger the amount of the Mn atoms incorporated in the Sb layers, the stronger the tendency to the interlayer FM coupling is \cite{Murakami.prb2019, Liu.arxiv2020mst}. At lowest intermixing levels as well as some Mn deficiency, \MST, similarly to \MBT\, is experimentally found to be an interlayer antiferromagnet, in which FM Mn layers of SLs are coupled antiparallel to each other below the N\'eel temperature of 19 K \cite{Liu.arxiv2020mst, Yan.PRB2019}. This is in agreement with our total energy calculations (Table~\ref{tab:bulks}), performed for the ideal crystal structure model. Importantly, we find that the ideal structure is energetically more favorable than those with Mn-Sb intermixing and therefore in what follows we consider the ordered structure of \MST.
We further find a positive magnetic anisotropy energy $E_a$
of 0.115 meV per formula unit, indicating the easy axis with
an out-of-plane orientation of the local magnetic moment (4.59 $\mu_B$), which is, again, in agreement with the experimental report
\cite{Yan.PRB2019}.

\begin{figure*}[t]
    \begin{center}
        \includegraphics[width=\textwidth]{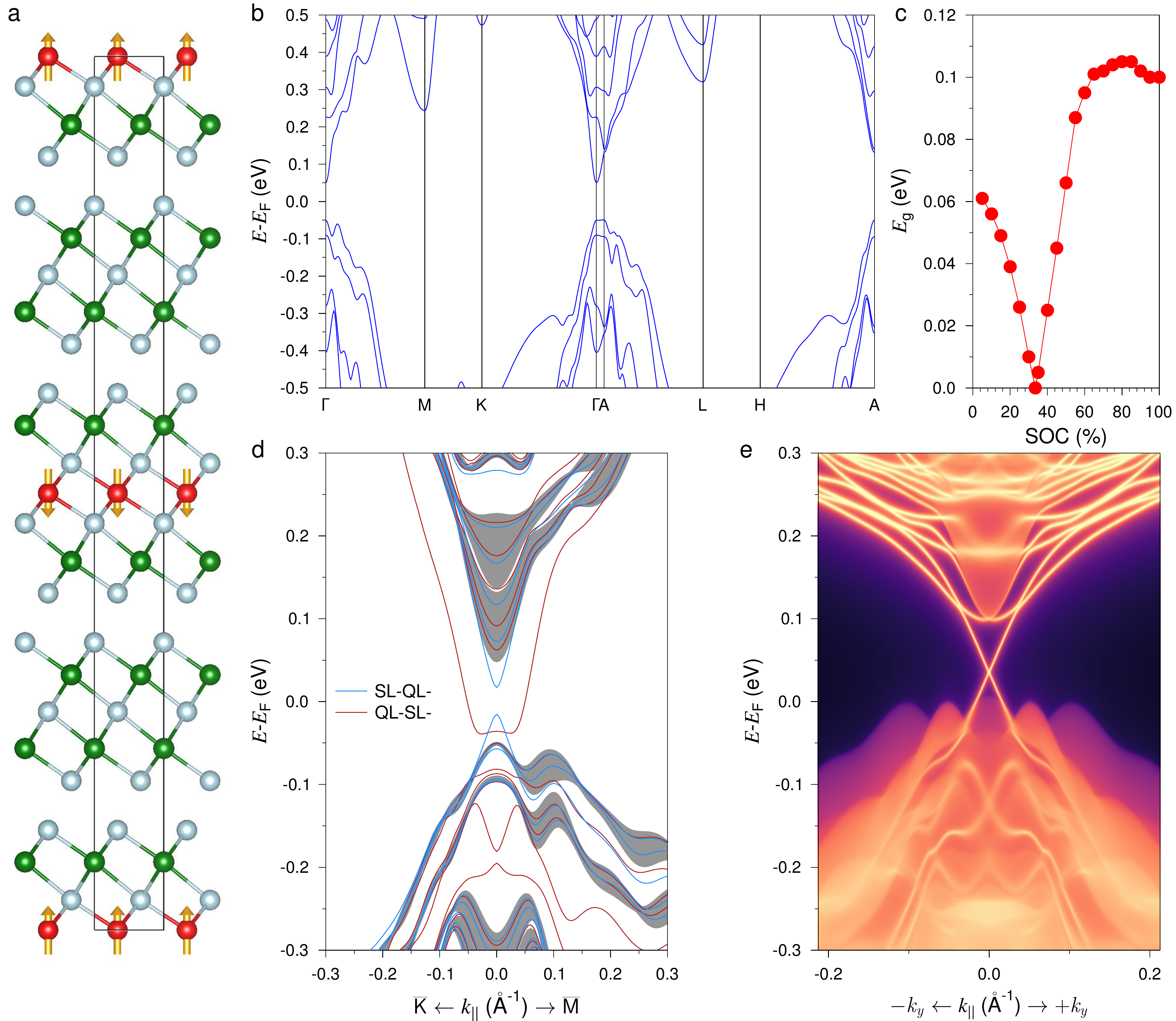}
    \end{center}
    \caption{{\bf Bulk and surface band structures of AFM \MSTST.} (a) Atomic structure of bulk \MSTST, with yellow, blue and green balls showing Mn, Te and Sb atoms, respectively. The interlayer AFM ground state is also reflected. (b) Bulk band spectrum calculated for the hexagonal unit cell. (c) The evolution of the bulk gap with the change of the SOC strength. (d) Surface band structure of \MSTST(0001) for the SL (blue) and QL (red) block terminations. (e) The tight-binding calculated electronic band structure of the $S$-preserving side surface.}
    \label{fig:147}
\end{figure*}

 \begin{table*}[t]
 \caption{Total energy differences of the AFM and FM configurations inside the Mn layer ($\Delta E_{||}=E_{NCAFM} - E_{FM}$) and between the neighboring layers ($\Delta E_{\perp}=E_{AFM} - E_{FM}$). In the former case, the AFM state is represented by the coplanar non-collinear configuration of local moments, in which the magnetizations of the three sublattices form angles of $120^\circ$ with respect to each other \cite{Eremeev.jac2017}. $E_{diff}=E_{in-plane} - E_{out-of-plane}$, is the total energy difference of the in-plane and out-of-plane (oop) directions of the magnetic moment, $E_{d-d}$ is the energy of the dipole-dipole interaction, $E_a=E_{diff}+E_{d-d}$ is the magnetic anisotropy energy, and $E_g$ is the bulk band gap.}
 \label{tab:bulks}
        \begin{center}
  \begin{tabular}{cccccccccc}
                \hline
                \hline
        &  $a_0$ & $c_0$ &   $m$   & $\Delta E_{||}$ & $\Delta E_{\perp}$ & $E_{diff}$ & $E_{d-d}$  &   $E_a$    & $E_g$\\
        &  (\AA) & (\AA) &($\mu_B$)&  (meV/Mn pair)  &   (meV/Mn pair)    & (meV/Mn at)&(meV/Mn at) &(meV/ Mn at)& (meV)\\
  \hline
  \hline
\MST    & 4.263  & 39.657&  4.59   &    17.30        & -1.30              & 0.24       &     -0.12  & 0.12  (oop)& 124 \\
\MSTST  & 4.265  & 23.265&  4.61   &    19.60        & -0.21              & 0.17       &     -0.12  & 0.05  (oop)& 100 \\
\MSTSTST& 4.269  & 99.563&  4.59   &    19.24        & -0.12              & 0.16       &     -0.12  & 0.04  (oop)& 91 \\
  \hline
  \hline
  \end{tabular}
        \end{center}
    \end{table*}

The combination of the trigonal structure and the interlayer AFM order makes \MST\, $S$-symmetric, $S$ being the combination of the time-reversal $\Theta$ and primitive-lattice translational symmetries $T_{1/2}$, $S=\Theta T_{1/2}$. The presence of this symmetry allows introducing $\mathbb{Z}_2$ classification of AFM insulators \cite{Mong.prb2010, Fang.prb2013}. To determine whether the system is insulating or not, bulk electronic structure has been studied next for the above discussed magnetic ground state. Our calculations reveal an insulating character of the spectrum, the fundamental band gap value being equal to 124 meV taking spin-orbit coupling (SOC) into account (Fig.~\ref{fig:124}b). The $\mathbb{Z}_2$ invariant can be calculated based on the occupied bands parities \cite{Mong.prb2010,Fang.prb2013}, in which way we find $\mathbb{Z}_2=1$ for \MST, meaning that it is an AFMTI. The latter means that the bulk band gap of the material should be inverted, which is indeed confirmed by performing the density of states (DOS) calculations lowering the SOC constant $\lambda$ stepwise from its natural value $\lambda_0$ to $\lambda=0.35\lambda_0$ (see Fig.~\ref{fig:124}c). As can be seen, the system demonstrates the topological phase transition, i.e. passes through the zero gap state, when the SOC strength is decreased to about 0.7$\lambda_0$. Recent DFT studies \cite{Zhang.prl2019, Chen.ncomms2019, Zhou.prb2020, Lei.arxiv2020, Liu.arxiv2020mst} report, however, that \MST\, is topologically trivial, which is in contrast to our results.
As we discuss in detail in the Supplementary Information, the crucial factor causing this difference is the \MST\, crystal structure, which in our case is fully optimized taking the van der Waals interactions into account (see Methods section). To provide further support to our claim of the AFMTI state in \MST, apart from the projector augmented wave method (PAW) calculations using VASP, we have performed DFT calculations using the full-potential linearized augmented plane-wave method (FLEUR code) as well as the Green's functions method (HUTSEPOT code). These calculations predict the band-inverted AFMTI phase in \MST\, as well (Supplementary Information Figures S1 and S2).

The bulk edge correspondence dictates that the inverted bulk band gap gives rise to a 2D spin-polarized state at the material's surface, i.e. the topological surface state (TSS). In an AFMTI case, this state may either be gapless or not, depending on whether the corresponding surface preserves the $S$ symmetry. Namely, the breakdown of this symmetry lifts the Dirac point (DP) protection and even an infinitesimal magnetization component perpendicular to the surface will cause the DP gap opening \cite{Mong.prb2010,Fang.prb2013, Otrokov.Nature, Zhang.prl2019, Li.sciadv2019}. To
check whether \MST\, shows such a behavior, we have calculated the electronic structures of the $S$-breaking [(0001)] and $S$-preserving [($10 \bar 1 1$)] surfaces. In the latter case, since the computational cell is very large, the calculations have been done using the \emph{ab-initio}-based tight-binding method, which is more cost-effective computationally. In Fig.~\ref{fig:124}d one can see that the spectrum of the $S$-breaking (0001) surface is indeed gapped, with the DP gap being equal to 14.8 meV. The spin texture of the gapped Dirac state (Fig.~\ref{fig:124}e) changes from a helical far from the $\bar\Gamma$ point to the hedgehog one in the vicinity of the gap with a purely out-of-plane spin alignment at $\bar\Gamma$, positive and negative for the lower and upper branches, respectively. The $S$-preserving ($10 \bar 1 1$) surface features a gapless Dirac cone (Fig.~\ref{fig:124}f), as expected for an AFMTI. These results constitute another proof of the AFMTI phase in \MST.

The prediction of the AFMTI state in \MST\, implies that there exists a number of related intrinsically magnetic insulators that should be topologically nontrivial.
Indeed, as a rule, compounds with a formula $XY_2Z_4$ ($X$=Ge, Sn, Pb; $Y$ = Sb, Bi and $Z$ = Se, Te), give rise to a series
of the  $(XY_2Z_4) \cdot (Y_2Z_3)_n$ form that retain a nontrivial topology \cite{Eremeev_NatComm2012,Silkin.jetpl2012,Papagno.acsn2016}.
Let us therefore consider a hypothetical MnSb$_4$Te$_7$ compound, whose crystal structure is formed by alternating quintuple layers (QLs) of \ST\, and SLs of \MST\, (space group $P \bar 3 m1$; Fig.~\ref{fig:147}a). Our total-energy calculations reveal that each SL block of \MSTST\ is ordered ferromagnetically, while the interlayer exchange coupling is antiferromagnetic (see Table~\ref{tab:bulks} and Fig.~\ref{fig:147}a). Further calculations show that the system has an easy axis pointing out of the Mn plane, but the magnetic anisotropy energy, $E_a$, appears to be significantly lower than that in \MST: 0.05 meV. The reduction of $E_a$ arises due to the reduced $E_{diff}$ value, while the dipolar contribution stays the same because both the local magnetic moment and in-plane lattice parameter are practically equal to those  in \MST, see Table~\ref{tab:bulks}. It is reasonable to suggest that, similarly to how it happens in the (\MBT)$\cdot$(\BT)$_n$ series \cite{Aliev.jac2019,Klimovskikh_npjQM2020,Hu.NatComm2020,Wu.sciadv2019}, the critical temperature of the magnetic ordering will drop roughly twice when going from \MST\, to \MSTST. This is due to the weakening of both the interlayer exchange coupling and magnetic anisotropy energy of the system (Table~\ref{tab:bulks}).

Like in the \MST\, case, the combination of the $P \bar 3 m1$-structure with the interlayer AFM ordering in \MSTST\, meets the $S$-symmetry. According to the bulk electronic structure calculations, \MSTST\ features an inverted bulk band gap of 0.1 eV (Fig.~\ref{fig:147}b,c), while the calculated $\mathbb{Z}_2$ value of 1 confirms nontrivial topology of the system, which similarly to \MST\ appears to be an AFMTI.

An important difference between the \MST\, and \MSTST\, systems is a possibility of having two terminations at the \MSTST(0001) surface: the QL and SL ones. Consistently with the bulk edge correspondence principle, there is a TSS at both these terminations (Fig.~\ref{fig:147}d), but the dispersions of the two TSSs are strongly different. The TSS of the
SL termination is located within the fundamental band gap and shows the DP gap of 32.5 meV, which is due to the $S$-symmetry breaking at the \MSTST(0001) surface and the out-of-plane magnetization of the Mn layer. The dispersion seen at the QL termination appears to be much more complex, owing to more complex localization of the QL-terminated TSS in TIs with alternating QL/SL structure \cite{Eremeev_NatComm2012,Grimaldi_PRB2020} which shows relocation of the state from surface QL to subsurface SL block in the vicinity of the $\bar\Gamma$ point. Such a behavior usually results in deviation of the TSS dispersion from linear near $\bar\Gamma$ and even Dirac state can demonstrate strongly pronounced kink-like dispersion \cite{Grimaldi_PRB2020}. In MnSb$_4$Te$_7$, when the magnetic moments on Mn atoms are artificially constrained to zero (Fig.~S3 in Supplementary Information), the TSS at the QL termination also demonstrates strong deviation from linear dispersion at $k_\| \approx 0.035$~\AA$^{-1}$. At the QL termination of AFM MnSb$_4$Te$_7$(0001) the Dirac state has magnetic gap of 46.3 meV at $\bar\Gamma$, however, the lower branch of the state appears at -0.08 eV, between the magnetically split subbands of the upper bulk valence band (Fig.~\ref{fig:147}(d)).
In contrast, the $S$-preserving side surface hosts a single gapless TSS  (Fig.~\ref{fig:147}e).

\begin{figure*}[t]
    \begin{center}
        \includegraphics[width=\textwidth]{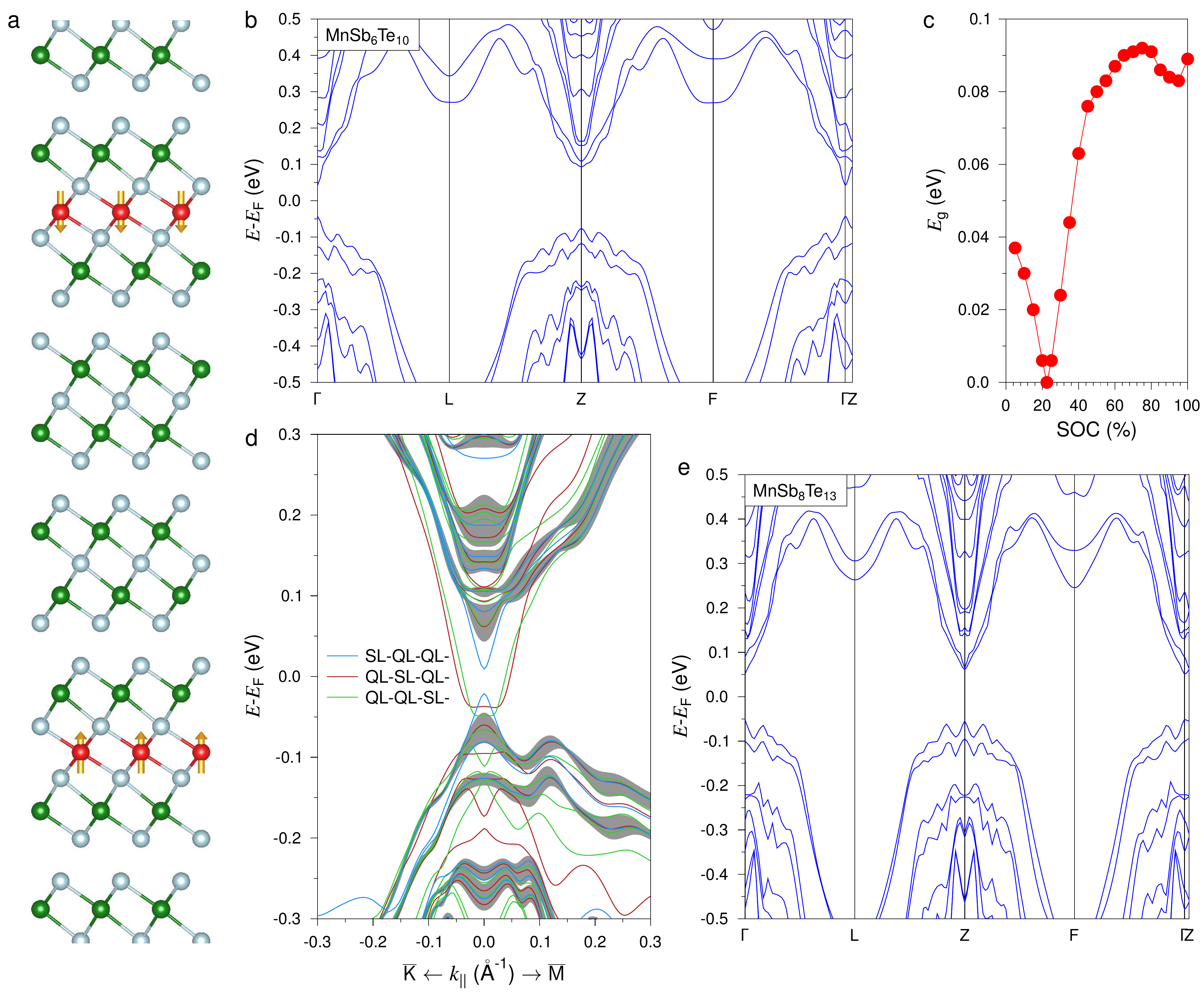}
    \end{center}
    \caption{{\bf Bulk and surface band structures of AFM \MSTSTST.} (a) Atomic structure of bulk \MSTSTST, with yellow, blue and green balls showing Mn, Te and Sb atoms, respectively. The interlayer AFM ground state is also reflected. (b) Bulk electronic structure calculated for the rhombohedral unit cell. (c) The evolution of the bulk band gap $E_g$ with the change of the SOC strength. (d) Surface band structure of \MSTSTST(0001) for the SL-QL-QL- (blue), QL-SL-QL- (red), and QL-QL-SL- (green) terminations. Gray areas correspond to the bulk band structure projected onto the surface Brillouin zone. (e) Bulk spectrum of MnSb$_8$Te$_{13}$ in the AFM state.}
    \label{fig:1610}
\end{figure*}

Another potential member of the (\MST)$\cdot$(\ST)$_n$ family is \MSTSTST. In this compound, adopting a trigonal structure with the $R\bar 3m$ space group, each \MST\, SL is separted by two \ST\, QLs (Fig.~\ref{fig:1610}a), leading to even weaker exchange interaction between SLs, which, nevertheless, stays antiferromangetic (Table~\ref{tab:bulks}). The bulk spectrum has a direct $\Gamma$-point gap of 91 meV (Fig.~\ref{fig:1610}b), which is inverted as confirmed by its dependence on the SOC strength (Fig.~\ref{fig:1610}c). Our $\mathbb{Z}_2$ invariant calculations (\MSTSTST\, is $S$-symmetric as well) yield a value of 1, meaning that it is an AFMTI, too.

Sharing the same crystal structure with MnBi$_6$Te$_{10}$ \cite{Klimovskikh_npjQM2020} and non-magnetic PbBi$_6$Te$_{10}$ \cite{Eremeev_NatComm2012, Papagno.acsn2016}, \MSTSTST\ should have three possible surface terminations, QL-QL-SL-, QL-SL-QL- and SL-QL-QL-,  showing different dispersions of the TSS (Fig.~\ref{fig:1610}d). The spectra of the SL-QL-QL- and QL-SL-QL-terminated surface of \MSTSTST\ in general are similar to those for SL-QL- and QL-SL- terminations in MnSb$_4$Te$_7$, c.f. Fig.~\ref{fig:147}d and Fig.~\ref{fig:1610}d. The TSS dispersion at the double-QL termination, QL-QL-SL-, is similar to that of the QL-SL-QL- termination as well.

A common feature of \MSTST\, and \MSTSTST\, is a weak interlayer exchange coupling. As it can be seen from Table~\ref{tab:bulks}, the total-energy difference between the interlayer AFM and FM structures, $\Delta E_{\perp}= E_{AFM} - E_{FM}$, in \MSTST\, (\MSTSTST) is more than 6 (10) times smaller than that in \MST, indicating a dramatic drop in the interlayer exchange coupling strength due to insertion of one (two) \ST\, QL block(s) between each pair of \MST\, SLs. With further increase in number of the \ST\ QL blocks to three and formation of the MnSb$_8$Te$_{13}$ compound the interlayer exchange coupling between ferromagnetic SLs practically vanishes. This leads to a situation where below the Curie temperature of the \MST\, SLs their magnetizations are randomly oriented either parallel or antiparallel to the [0001] direction. If calculated in the imposed interlayer AFM state, MnSb$_8$Te$_{13}$ shows  an inverted bulk band gap of about 100 meV (Fig.~\ref{fig:1610}e).

We envision, that similarly to (\MBT)$\cdot$(\BT)$_n$ \cite{Jahangirli.jvst2019, Klimovskikh_npjQM2020}, the (\MST)$\cdot$(\ST)$_n$ homologous series can contain members with $n>3$, i.e. MnSb$_{10}$Te$_{16}$, MnSb$_{12}$Te$_{19}$, etc., that are going to be topologically-nontrivial as well. This is because adding more \ST\, QLs is expected to make the band gap inversion stronger, since the material becomes more Sb$_2$Te$_3$-like. Thus, if an $n \geq 3$ compound is cooled below the critical temperature of the SL block FM ordering, a special phase is formed, in which a nontrivial topological order coexists with a partial magnetic disorder in the system of the decoupled 2D ferromagnets, whose magnetizations point randomly along the third direction.

\begin{figure*}[t]
    \begin{center}
    \includegraphics[width=\textwidth]{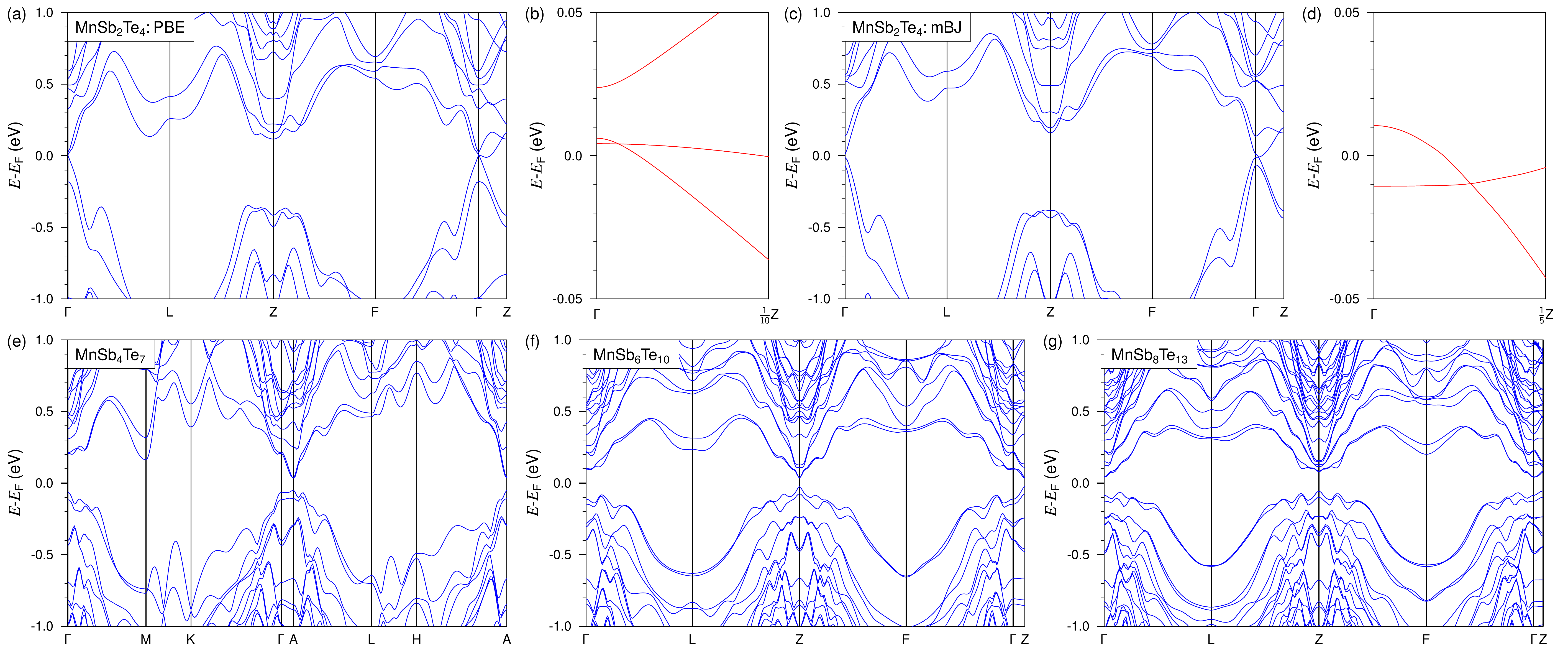}
    \end{center}
    \caption{{\bf Bulk band structure of ferromagnetic phase of \MST\, family compounds.} (a) Spectrum of \MST\ as calculated within GGA-PBE$+U$; (b) Magnified view of the \MST\ spectrum near the Fermi level in the vicinity of the $\Gamma$ point; (c) and (d) the same as in panels (a) and (b) but calculated with mBJ functional with Hubbard $U$; Bulk spectra of ferromagnetic MnSb$_4$Te$_7$ (e), MnSb$_6$Te$_{10}$ (f), and MnSb$_8$Te$_{13}$ (g) calculated along high symmetry directions of corresponding Brillouin zones (hexagonal or rhombohedral) within GGA-PBE$+U$.}
    \label{fig_FM}
\end{figure*}

The relatively weak interlayer exchange coupling in \MST, \MSTST, and \MSTSTST\, or its absence for $n \geq 3$ make possible driving them into a globally FM phase by applying an external magnetic field along the $c$ axis. Such a change in a magnetic state may cause change in the topology. In order to identify the topological phases of the (\MST)$\cdot$(\ST)$_n$ family compounds in FM state we calculated the strong topological invariant $\mathbb{Z}_4$ (see Methods section). $\mathbb{Z}_4$ = 1 or 3 correspond to Weyl semimetal phases, where an odd number of Weyl points exist in half of the BZ. $\mathbb{Z}_4$ = 2 corresponds to the AXI, having the quantized topological magnetoelectric response in the bulk and chiral hinge modes.

Similarly to \MBT, where in FM state a Weyl semimetal phase was predicted \cite{Li.sciadv2019,Zhang.prl2019}, ferromagnetic \MST, according to $\mathbb{Z}_4$ calculations, is a Weyl semimetal, too ($\mathbb{Z}_4=1$), in agreement with Refs. \cite{Murakami.prb2019, Lei.arxiv2020}. Namely, the GGA$+U$ bulk spectrum (Fig.~\ref{fig_FM}a,b) is gapless and shows a band crossing near the Fermi level, forming the Weyl points ($k_{W}$) at $\pm$0.003~\AA$^{-1}$\ along the $\Gamma$Z direction just above the Fermi level. However, since the inversion of the bands at $\Gamma$ is only 2 meV, the topological phase can be sensitive to the choice of the exchange-correlation functional. As can be seen in Figs.~\ref{fig_FM}c,d, applying the modified Becke-Johnson (mBJ) semilocal exchange potential \cite{mBJ1,mBJ2} retains the Weyl phase, providing larger $\Gamma$ gap (20 meV) and a larger $k_{W}$=$\pm$0.01~\AA$^{-1}$\ as well as shifting the Weyl points slightly below the Fermi level.

The bulk spectra of MnSb$_4$Te$_7$ (Fig.~\ref{fig_FM}e) and MnSb$_6$Te$_{10}$ (Fig.~\ref{fig_FM}f) in the FM phase demonstrate direct band gaps at the BZ boundaries with magnitudes of 84 and 58 meV, respectively, whereas the  MnSb$_8$Te$_{13}$ spectrum shows indirect $\Gamma$-Z gap of 96.5 meV (Fig.~\ref{fig_FM}g). At the same time, all these spectra demonstrate band inversion of the Sb and Te $p_z$ orbitals at the $\Gamma$ point. Calculations of the topological indices for ferromagnetic phase of the (MnSb$_2$Te$_4$)$\cdot$(Sb$_2$Te$_3$)$_n$ compounds with $n$=1, 2, and 3 gives $\mathbb{Z}_4$ = 2 and hence all of them are in the FM AXI state, which can be expected for $n > 3$ as well.

Now we show that in the thin-film limit not only the magnetic, but also the electronic and topological properties of \MST, are strongly thickness dependent. The spectrum of a single \MST\ SL demonstrates the indirect gap of 419.7 meV, which is significantly larger than that in the \MBT\ SL ($\sim\SI{0.32}{\eV}$)\cite{Otrokov.prl2019,Gong_CPL2019}. The direct $\bar\Gamma$-gap amounts to 811.4 meV. The Chern number calculations reveal a $C=0$ state, the system being a topologically trivial ferromagnet (Table~\ref{tab:topo}).
Upon increasing thickness up to 2, 3 and 4 SLs the interlayer AFM order (compensated or uncompensated for even and odd SL thicknesses, respectively) sets in, but $C=0$ again for all of them. For these systems, we find the $\bar\Gamma$-point band gaps of 210.2, 69.5, and 65.8 meV, respectively (see Table~\ref{tab:topo}). However, if calculated in the artificial FM phase (that could be achieved by applying external magnetic field), the Chern number appears to be equal to $-1$ in case of the 4-SL-thick film (for 2 and 3 SLs $C$ is still zero). This means that 4 SLs of \MST\ is a Chern insulator that is expected to show the quantum Hall effect in the external magnetic field but without Landau levels, as it has been recently reported for \MBT\, flakes \cite{Liu.NatMat2020, Deng.Sci2020, Ge.NSR2020}. Reversing the magnetization of the FM 4-SL-thick \MST\ film yields the $C=+1$ Chern insulator state, thus proving a so-called zero plateau quantum Hall behavior in the compensated AFM state of this film, characteristic of the AFM AXI \cite{Otrokov.prl2019, Li.sciadv2019, Liu.NatMat2020}.

 \begin{table}[t]
 \caption{ Thickness dependence of the MnSb$_2$Te$_4$ films topology and the $\bar\Gamma$ band gap. } \label{tab:topo}
 \begin{center}
 \begin{tabular}{ccc}
                \hline \hline
                Thickness (SL)    & Topology    & $\bar\Gamma$-gap (meV)\\
                \hline \hline
                1                 & Trivial            & 811.4 \\
                2                 & Trivial            & 210.2 \\
                3                 & Trivial            &  69.5 \\
                4                 & AFM AXI              &  65.8 \\
                5                 & Trivial            &  18.6 \\
                6                 & AFM AXI              &  30.4 \\
                7                 & QAHI                &  4.8 \\
                8                 & AFM AXI              &  14.8 \\
                \hline
                $\propto$ (bulk)  & 3D AFMTI          &  124.0 \\
                \hline \hline
            \end{tabular}
        \end{center}
    \end{table}

At a thickness of 5 SLs \MST\, appears to be in the trivial insulator phase again, the band gap being equal to 18.6 meV. Similarly to the 4-SL-thick film case, we also predict the intrinsic AFM AXI states for the 6- and 8-SL-thick films (meaning that in the FM state they are expected to show the quantum Hall effect without Landau levels), while the 7-SL-thick \MST\ film is in the QAH phase with a gap of 4.8 meV (Table~\ref{tab:topo}). Thus, like \MBT\ in the thin film limit \cite{Otrokov.prl2019, Li.sciadv2019} the \MST\ films also demonstrate alternating AXI/QAHI topological phases, however, owing to a weaker SOC in \MST, they emerge at larger thicknesses than in the \MBT\, case.

We have further constructed various \MST/(\ST)$_n$/\MST\, sandwiches with different number of the \ST\, QLs, $n = 1 - 3$. While for $n=1$ and 2 we find topologically trivial situation for both parallel and antiparallel alignments of the SLs magnetizations, the $n=3$ system appears to be in the AFM AXI and Chern insulator phases for the AFM and FM states, respectively. Similar behavior is also expected for $n=4$ and 5. Note that \MST/(\ST)$_n$/\MST\, can in principle be obtained by careful exfoliation of the thin flakes from the (\MST)$\cdot$(\ST)$_n$ single crystals.

\section*{Conclusions}

Using state-of-the-art \emph{ab initio} calculations we have predicted the existence of a new family of magnetic topological insulators of the\, van\, der\, Waals\, layered\, compounds\,  (\MST)$\cdot$(\ST)$_n$. In their groundstate, the \MST\, ($n=0$), \MSTST\, ($n=1$), and \MSTSTST\, ($n=2$) materials appear to be antiferromagnetic topological insulators. For $n \geq 3$ (i.e. from MnSb$_8$Te$_{13}$ and on), a
special magnetic topological insulator phase is formed in which, below critical temperature, the magnetizations of the 2D ferromagnetically ordered Mn layers of the \MST\, building blocks are disordered along the [0001] direction. We have further shown that imposing the overall ferromagnetic state (which can be achieved by applying external magnetic field) drives \MST\, into a Weyl semimetal phase, while the compounds with higher $n$ become 3D ferromagnetic axion insulators. Finally, in the 2D limit, (\MST)$\cdot$(\ST)$_n$ films show a variety of the topologically-nontrivial states, being among them intrinsic axion and quantum anomalous Hall insulators, as well as quantum Hall state, that could be achieved under external magnetic field, but without Landau levels. Our results convincingly show that \MST\, is not trivial as it has been concluded by the previous \emph{ab initio} works and we actually demonstrate the reason why those studies found it to be trivial. Obviously, the (\MST)$\cdot$(\ST)$_n$ family represents a fertile ground for realizing multiple and tunable exotic states and the experimental realization of this family would represent a great advance in the research of the magnetic topological matter.

\emph{Note.} Very recently, the first member of the (MnSb$_2$Te$_4$)$\cdot$(Sb$_2$Te$_3$)$_n$ series, i.e. \MST, has been proved in the experiment as a magnetic topological insulator \cite{Wimmer.arxiv2020}.

\section*{Methods}
\subsection*{Electronic structure and total-energy calculations}

Electronic structure calculations were carried out within the density functional theory using the projector augmented-wave (PAW) method \cite{Blochl.prb1994} as implemented in the VASP code \cite{vasp1, vasp2}. The exchange-correlation energy was treated using the
generalized gradient approximation \cite{Perdew.prl1996}. The Hamiltonian contained scalar relativistic corrections and the spin-orbit coupling was taken into account by the second variation method \cite{Koelling.jpc1977}. The energy cutoff for the plane-wave expansion was set to 270 eV. The crystal structure of each compound was fully optimized to find the equilibrium lattice parameters, cell volumes, $c/a$ ratios as well as atomic positions. At that, a conjugate-gradient algorithm was used. In order to describe the van der Waals interactions we made use of the DFT-D2 \cite{Grimme.jcc2006} and the DFT-D3 \cite{Grimme.jcp2010, Grimme.jcc2011} approaches, which gave similar results. The atomic coordinates were relaxed using a force tolerance criterion for convergence of 0.01 eV/{\AA}. Spin-orbit coupling was always included when performing relaxations.

The Mn $3d$-states were treated employing the GGA$+U$ approach \cite{Anisimov1991} within the Dudarev scheme \cite{Dudarev.prb1998}. The $U_\text{eff}=U-J$ value for the Mn 3$d$-states was chosen to be equal to 5.34~eV, as in previous works \cite{Otrokov.jetpl2017,Otrokov.2dmat2017,Hirahara.nl2017, Eremeev.nl2018, Otrokov.Nature, Otrokov.prl2019, Klimovskikh_npjQM2020, Hirahara.ncomms2020}.
Further extensive testing was performed for \MST, \MSTST\, and \MSTSTST\, systems to check the stability of the results against $U_\text{eff}$ variation. Namely, the bulk crystal structure was fully optimized
for $U_\text{eff}=3$ eV and then the magnetic
ordering and electronic structure were studied.  It was found that neither the magnetic ground state nor the topological class change upon such a variation of $U_\text{eff}$ and crystal structure. The magnetic anisotropies of (\MST)$\cdot$(\ST)$_n$ were found to be stable against this variation as well.

The bulk magnetic ordering was studied using total-energy calculations, performed for the FM and two different AFM states. Namely, we considered an interlayer AFM state and a noncollinear intralayer AFM state, in which three spin sublattices form angles of 120$^\circ$ with respect to each other \cite{Eremeev.jac2017}. To model the FM and interlayer AFM structures in (\MST)$\cdot$(\ST)$_n$, we used rhombohedral cells for $n=0,2$, and 3, while a hexagonal cell was chosen for $n=1$ (all cells contained the number of atoms corresponding to two formula units).
For the $n \le 2$ systems, the noncollinear intralayer AFM configuration was calculated using hexagonal bulk cells containing three atoms per layer [$(\sqrt{3}\times\sqrt{3})R30^\circ$ in-plane periodicity].
The magnetic anisotropy energy was calculated as explained in Ref. \cite{Otrokov.Nature}.

The (\MST)$\cdot$(\ST)$_n$ semi-infinite surfaces were simulated within a model of repeating films separated by a vacuum gap of a minimum of 10~\AA. The interlayer distances were optimized for the utmost SL or QL block of each surface.

Additional bulk electronic structure calculations for the AFM \MST\, (Supplementary Figure S1) were performed within the full-potential linearized augmented  plane  waves  (FLAPW) formalism  \cite{bib:wimmer81}  as implemented  in FLEUR \cite{bib:fleur}. We took the GGA+$U$ approach under the fully localised limit with the same $U_{\rm eff}$=5.34 eV as we use in VASP calculations. The core states were treated fully relativistically, while the valence states were computed in a scalar relativistic approximation with taking into account the spin-orbit coupling. The muffin-tin radii for Mn, and Te were chosen to be 2.70 a.u., while for Sb we used 2.87 a.u. Inside each muffin-tin sphere, the basis functions were expanded in spherical harmonics with angular momentum up to $l_{\rm max}$=10, and the electron density and potential were expanded up to $l_{\rm max}$=8. The crystal wave functions were expanded into augmented plane waves with a cutoff of $k_{\rm max}$=3.7 $a.u.^{-1}$, corresponding approximately to the 180 basis LAPW functions per atom.

The bulk electronic structure of the AFM \MST\, was also calculated using the HUTSEPOT code based on the Green's function method within the multiple scattering theory \cite{Geilhufe2015} (Supplementary Figure S2). To describe both localization and interaction of the Mn $3d$ orbitals, the Hubbard $U_\text{eff}$ values between 3 and 5 eV were employed using the GGA+$U$ approach \cite{Anisimov1991}.

Ab-initio-based tight-binding calculations were performed using the VASP package with the Wannier90 interface \cite{marzari_vanderbilt1997,Mostofi2008}. The Wannier basis chosen consisted of six spinor $p$-type orbitals of Sb and Te. The surface electronic band structure was calculated within the semi-infinite medium Green’s function approach \cite{Sancho:85,Henk:93}.

\subsection*{Topological invariants calculations}
The Chern numbers were calculated using Z2Pack \cite{Soluyanov.prb2011,Gresch.prb2017}.  The $\mathbb{Z}_2$ invariant for the 3D AFMTIs was calculated using the expression derived in Ref.~[\onlinecite{Fang.prb2013}]. When spatial inversion symmetry is present in the system, as it is in the \MST-family compounds, the following $\mathbb{Z}_2$ invariant $\lambda_0$ can be defined:
$$
(-1)^{\lambda_0} = \prod_{{\mathbf k}_{inv} \in \mathrm{B-TRIM}, n \in occ/2}{\zeta_n({\mathbf k}_{inv})},
$$
where $\zeta_n$ is the parity of the \textit{n}th occupied band and the multiplication over $occ$/2 means only one state in a Kramers' pair is chosen. See Ref.~[\onlinecite{Fang.prb2013}] for further details.

Calculations of $\mathbb{Z}_4$ topological invariant were done according to Refs.~\cite{XuZ4} as
$$
\mathbb{Z}_4 = \sum_{\alpha=1}^8\sum_{n=1}^{n_{occ}}\frac{1+\zeta_n(\Lambda_{\alpha})}{2} \mod 4,
$$
where $\Lambda_{\alpha}$ are the eight inversion-invariant momenta, $n$ is the band index, $n_{occ}$ is the total number of electrons, and $\zeta_n(\Lambda_{\alpha})$ is the parity of the $n$-th band at $\Lambda_{\alpha}$.

\bibliographystyle{ScienceAdvances}

\end{document}